\documentclass{icya}
\begin{document}

\title[ICYA2009]{Non-thermal images of SN 1006: from radio to $\gamma$-rays}

\author[V.Beshley, IAPMM]{Vasyl Beshley\thanks{Beshley.vasyl@gmail.com, Institute for Applied Problems in Mechanics and Mathematics} }


\newcommand{\DTP}{{\sffamily\bfseries Desktop Publishing}}
\newcommand{\memoir}{{\sffamily\bfseries memoir}\ }
\newcommand{\texlive}{\TeX live\ }
\newcommand{\bibtex}{Bib\TeX\ }

\iabstract{%
Supernova remnants (SNRs) are believed to be the main sources of galactic cosmic rays. Discovery of the non-thermal component in X-ray spectrum of SN 1006 in 1995 and detection of a number of SNRs by H.E.S.S. strengthen the investigation of SNRs. SN 1006 remains to be one of the most interesting objects for high-energy astrophysics. Electrons accelerated by the shock are the source of the non-thermal radiation in radio, X-rays (via synchrotron emission) and $\gamma$-rays (via inverse-Compton process). Experimental images of SN 1006 are known in all these bands, including the very-high energy $\gamma$-ray range. An important task is therefore to model the distribution of the surface brightness in SNRs. We develop a method for synthesis of SNR maps due to the non-thermal radiation of electrons in radio, X-rays and $\gamma$-rays. In particular, the method takes into account the injection of particles and the behaviour of magnetic field at shocks with different obliquities as well as the radiation losses of electrons downstream of the shock. The method is used to model images of SN 1006. The images well correlate with the observations in radio and X-ray but not in $\gamma$-ray range. For coincide in $\gamma$-ray range it is needed to increases parameters $\eta$ and $E_{f}$.}

\maketitle

\section{Introduction}
\indent \indent It is common believe that SNRs are the main sources of cosmic rays at least up to the knee at $3\times10^{15}$ eV and even beyond \citep{bervolk}, \citep{blandford}. Electrons accelerated at the shock of SNRs radiate in wide range from radio to $\gamma$-rays. The SN1006 is the prime example of such an object. This object is a type of Ia SNR located at a distance of 2.2 kpc \citep{winkler}. 

The remnant of SN 1006 was first identified in the radio \citep{gurdner}. In 1982, the first paper went describing were published about describe the first radio images of SN 1006 and Tycho \citep{reynolds82}. 

Synchrotron emission from accelerated electrons is detected up to the X-rays in SN 1006 \citep{rothenflug}. In 1995, the first papers went out about domination of non-thermal X-ray component in some regions of SN 1006 \citep{koyama}. 

Investigation of the high energy $\gamma$-rays was the next step. The first observations in this range are done by the CANGAROO telescope \citep{tanimori}. The first images of SN 1006 as a TeV source was reported by H.E.S.S. team in 2008.

It is suitable to model this object because it has the simple symmetrical distribution of surface brightness in all ranges. Almost spherical form of SN 1006 provide evidences that it develops in a uniform interstellar medium (ISM). Therefore, we model SN 1006 and synthesize it is images in different bands as SNR in uniform ISM and uniform interstellar magnetic field (ISMF).

\begin{figure*}[!ht]
\centering
	\includegraphics[width=11truecm]{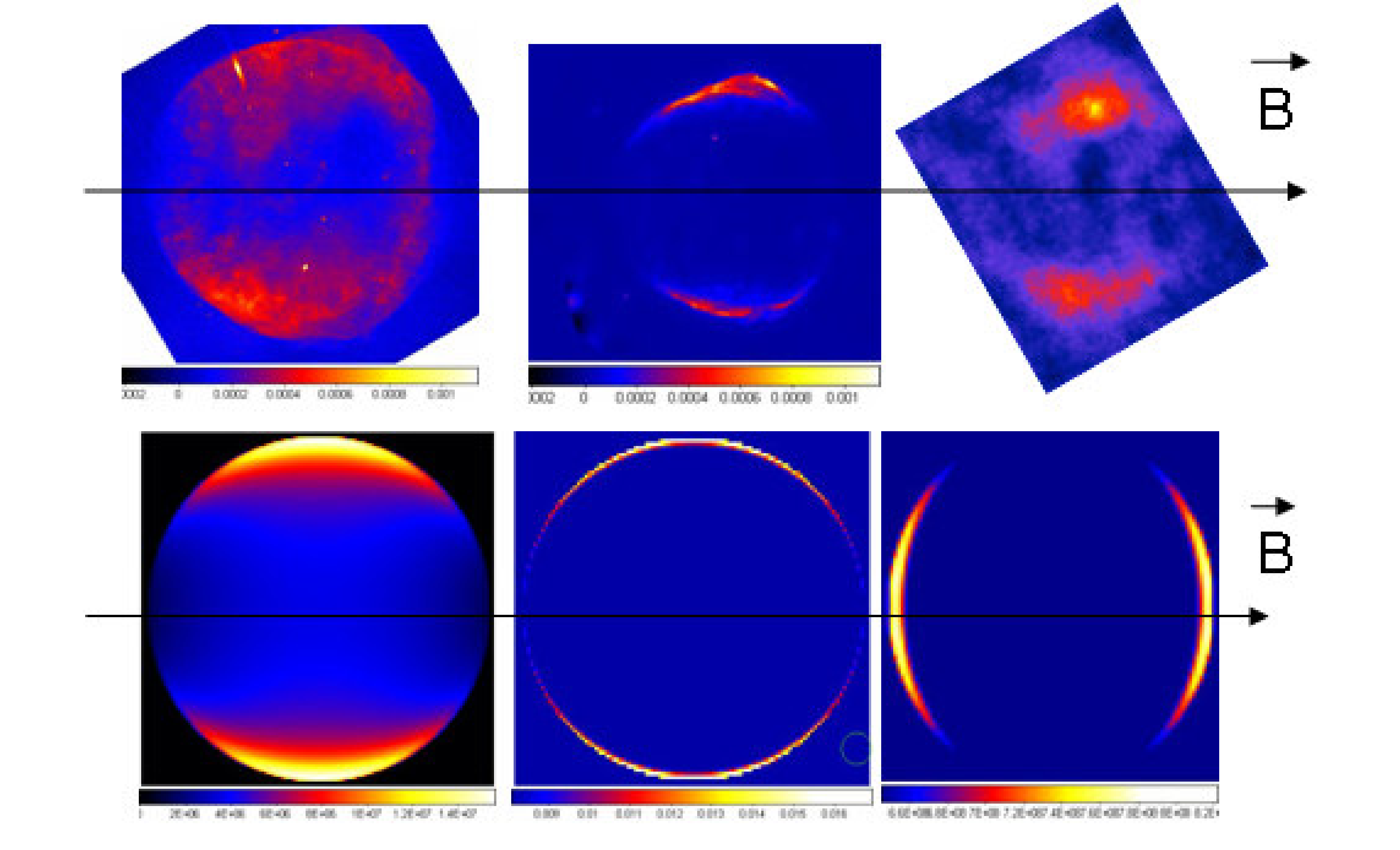}
\caption[images]{Observational(upper panel) and theoretical(lower panel) images of SN 1006 in different bands, from left to right; radio (1517 MHz); X-rays (3.5;2-4.5 $\times 10^{3}$ eV); $\gamma$-rays (1;0.5-10 $\times 10^{12}$ eV)}
\label{fig}
\end{figure*}

\section{Description of The Model}

\indent \indent SN 1006 is on the adiabatic stage of evolution. Therefore, for description of hydrodynamics we use the Sedov self-similar solutions \citep{sedov}. The calculations of the magnetic field in the SNR interier and on the shock, are bathed on the model of \citet{reynolds}. We use the classic model of kinetics of the charged particles. The spectrum of relativistic electrons is power-law with the exponential cut-off: 
\begin{equation}  
 N(E)dE=KE^{-s}\exp\left(-{E}/{E_{max}}\right)dE
 \label{sp-exp}
\end{equation}
where $s$ is a spectral index and $E_{max}$ the maximum energy of electrons. Distribution of spectrum of electrons is downstream of the shock is described in paper \citep{reynolds} and generalized in \citep{pet-besh2}.

It is also needed to take into account the azimuth distribution of the maximal energy. It is not isotropic as it is given by observations \citep{rothenflug}. The Reynolds developed three models of maximum energy: 1)by the electron radiative losses (due to synchrotron and inverse Compton, loss-limited), 2) by the limited time of the acceleration (time-limited) and 3) by properties of microphysics (escape-limited). For description of distribution of maximum energy we use the second model (time-limited) with parameter $\eta$, defined as $\lambda=\eta r_{g}$, where $\lambda$ is the mean free path and $r_{g}$-gyroradius and $\eta$ the "gyrofactor".

\citet{reynolds} considered thee models for injection: quasi-parallel (qpar), isotropic (iso), quasi-perpendicular (qper). Morphology of SNR under quasi-perpendicular injection is similar to morphology for isotropic case \citep{orlando}. Experimental data reject quasi-perpendicular model \citep{petruk082}, if the SNR is assumed to be in uniform ISM. Therefore we consider only isotropic injection.

Spectral index. With the use of radio observation it is possible to find the value of spectral index. Spectral index in this case is $s=2.2$ \citep{rothenflug}.

Radiative losses. For the account of radiation losses of electrons we use the parameter $E_{f}$, $E_{f}=637\left(B^{2}tE_{max}\right)^{-1}$ is the reduced fiducial energy, where $t$ is the age of SNR, $B$ the magnetic field strength after parallel shock. This parameter is a measure of importable of radiative losses in modification of the electron spectrum \citep{reynolds}. If $0<E_{f}<1$ then the radiative losses are very important if $E_{f}>1$ radiative losses may be neglected. In our calculations $E_{f}=1$.

Other parameter is orientations of ISMF in respect to the observer ($\phi_{o}$). Aspect angel equals $70^{o}$ \citep{petruk082}.

In the present paper we consider two mechanisms of radiation of electrons. This is synchrotron \citep{pet-besh1}, \citep{pet-besh2} end inverse Compton(IC) \citep{petruk081}.

The surface brightness at an arbitrary point of SNR is calculated as integral along the line of sight, taking into account distributions of the magnetic field, density of electrons etc.

\section{Results}

\indent \indent Using results of observation we impose limitation on the parameters of model.

Maximum energy. Distribution of the maximum energy is taken from the observations of satellite XMM-Newton. For different azimuthal angle, $E_{max}$ is different. Observations agree well with time-limited model of \citet{reynolds} with $\eta=1$.

Using the classic model of kinetics of electrons the maps of surface brightness distribution  were built for three ranges: radio, X-ray and $\gamma$-ray. The model and observational images of SN 1006 are shown on Fig. \ref{fig}. The limbs of brightness in the simulated images coincide in radio and X-ray ranges. However we have no coincidence in $\gamma$-rays. The possible of such behaviour are discussed in \citep{pet-besh09}. The location of limbs, besides of azimuthal variations of injection efficiently and magnetic field compression, depends an a number of factors:
\begin{itemize}
\item azimuth distribution of the maximum energy of electrons given by parameter $\eta$
\item radiative losses controlled by the parameter $E_{f}$
\item non-uniformity of the interstellar medium
\item non-uniformity of the magnetic field.
\end{itemize}

With the growth of parameters $\eta$ and $E_{f}$, the maximal energy increases with the azimuthal angle and the radiation losses decrease. We expect that the values of these parameters $\eta$ and $E_{f}$ must change in a range 5-10 and 30-60 respectively, in order to provide coincidence of limbs also in $\gamma$-rays.

\section{Conclusions}

\indent \indent We considered the two models of emission of electrons: synchrotron (in radio, X-rays) and IC (in $\gamma$-rays) for the calculation of maps of surface brightness. Images are built in these three ranges (radio, X-ray and $\gamma$-ray) for SN 1006. The observational and theoretical images in X-rays and radio coincide well, but in $\gamma$-rays does not correlate. To match the images we should account for the following modifications in our  model:
\begin{itemize}
\item introduce non-uniform of ISM and ISMF
\item for consider $\eta$ in the range $ 30<\eta<60$
\item do such the role of smaller radiative losses ($E_{f}>30$).
\end{itemize}

\bibliography{beshley}
\end{document}